\documentclass[reprint, amsmath, amssymb, aps, superscriptaddress]{revtex4-1}
\usepackage{graphicx}
\usepackage{dcolumn}
\usepackage{bm}
\usepackage{epstopdf}

\newcommand{\ket}[1]{| #1 \rangle}
\newcommand{\bra}[1]{\langle #1 |}

\begin{document}

\preprint{APS/123-QED}

\title{Demonstration of the Jaynes-Cummings ladder with Rydberg-dressed atoms}

\author{Jongmin Lee}
\affiliation{Sandia National Laboratories, Albuquerque, NM 87185}
\author{Michael J. Martin}
\affiliation{Sandia National Laboratories, Albuquerque, NM 87185}
\author{Yuan-Yu Jau}
\affiliation{Sandia National Laboratories, Albuquerque, NM 87185}
\affiliation{Center for Quantum Information and Control (CQuIC), University of New Mexico, Albuquerque NM 87131}
\author{Tyler Keating}
\affiliation{Center for Quantum Information and Control (CQuIC), University of New Mexico, Albuquerque NM 87131}
\author{Ivan H. Deutsch}
\affiliation{Center for Quantum Information and Control (CQuIC), University of New Mexico, Albuquerque NM 87131}
\author{Grant W. Biedermann}\email{gbieder@sandia.gov}
\affiliation{Sandia National Laboratories, Albuquerque, NM 87185}
\affiliation{Center for Quantum Information and Control (CQuIC), University of New Mexico, Albuquerque NM 87131}

\date{\today}

\begin{abstract} 
We observe the nonlinearity of the Jaynes-Cummings (JC) ladder in the Autler-Townes spectroscopy of the hyperfine ground states for a Rydberg-dressed two-atom system. Here the role of the two-level system in the JC model is played by the presence or absence of a collective Rydberg excitation, and the bosonic mode manifests as the number $n$ of single atom spin flips, symmetrically distributed between the atoms. We measure the normal-mode splitting and $\sqrt{n}$ nonlinearity as a function of detuning and Rabi frequency, thereby experimentally establishing the isomorphism with the JC model.
\end{abstract}

\pacs{Valid PACS appear here}
\maketitle

The Jaynes-Cummings (JC) model~\cite{Cummings63} describes the interaction between a two-level atom and a single mode of the quantized electromagnetic field.  While originally introduced in the context of cavity quantum electrodynamics (QED)~\cite{Haroche85, Hinds92, Meystre92, CQED94} for single atoms~\cite{Rempe90, Kimble92, Feld94, Haroche96}, it applies also in solid state systems, where a qubit (encoded in two discrete energy levels) is strongly coupled to a cavity mode in the optical or microwave regime, as observed in quantum dots~\cite{Awschalom04, Langbein10, Wallraff13}, and most dramatically in superconducting circuits in which a phase, flux, or charge qubit is coupled to a quantized mode of a microwave cavity~\cite{Fink08, Finley12, Gross10}.  More generally, the JC model describes a spin-boson system where a qubit is coupled to a bosonic mode, e.g., it describes the dynamics of trapped atomic ions in which two internal atomic levels are coupled to a phononic mode of ion vibration~\cite{Wineland03}.

Given its simplicity, the nonlinear coupling of the JC model has been a staple of quantum optics for decades as a platform for quantum control~\cite{Haroche01a, Haroche01b}.  At its base is the spectrum of dressed states, the well known JC ladder, which exhibits nonlinear normal-mode splitting proportional to $\sqrt{n}$, for $n$ bosons coupled to the qubit on resonance.  This nonlinearity is responsible for the collapse and revival of Rabi oscillations~\cite{Eberly81, Knight82, Rempe87}, and the generation of nonclassical states, such as squeezed states~\cite{Vuletic10a, Vuletic10b, Thompson14, Thompson16, Kasevich16} and cat states. Spectroscopy of the JC ladder has been carried out in a single two-level atom in a high finesse cavity~\cite{Kimble92} and in a superconducting microwave circuit QED system~\cite{Fink08, Wallraff09}.  

In this letter we perform spectroscopy on a completely different instantiation of the JC model --- symmetric Rydberg-blockaded atomic ensembles~\cite{Molmer08, Molmer12, Beterov14, Keating16}. Here, the role of the qubit is played by the presence or absence of a collective Rydberg excitation, and the bosonic mode is the symmetric many-body ground state of an ensemble of $n$ identical atoms that can be coupled to the Rydberg level.  Similar to the cavity QED system, there is a $\sqrt{n}$ nonlinearity arising from the Rydberg blockade, and the symmetric coupling between a single Rydberg-excited atom and the collective ground-state of the atomic ensemble~\cite{Molmer02, Walker14, Evers14, Beterov14, Gross15, Ott15, Keating16}. In this system, the dressed states of the JC ladder are the laser-induced Rydberg-dressed states~\cite{Molmer02, Johnson10}.  The normal mode splitting is intimately related to the Autler-Townes splitting of the light shifted states~\cite{Killian16, Piotrowicz11}. The nonlinearity of the dressed-state spectrum was recently employed with two atoms to generate Bell states based on a spin-flip blockade~\cite{Jau15}. 

\begin{figure}[b!]
\includegraphics[width=1\columnwidth]{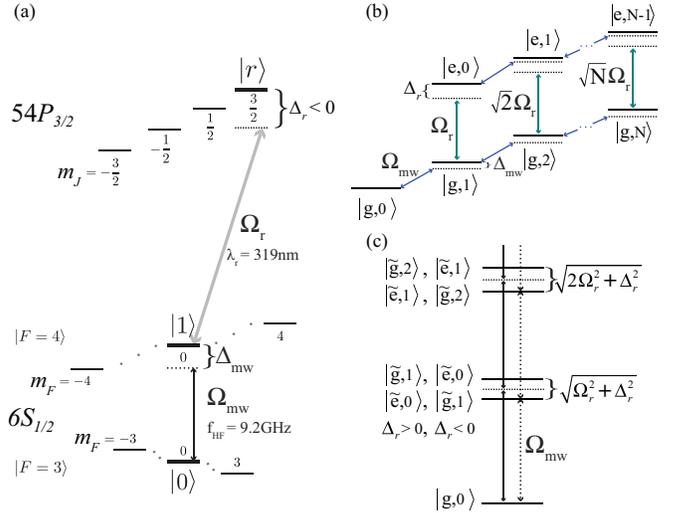}
\caption{(a) Energy level diagram of the $^{133}$Cs atom. The states are $|0\rangle = |6 S_{1/2}, F = 3, m_F = 0\rangle$ and $|1\rangle = |6S_{1/2}, F=4, m_F=0\rangle$, where $\Delta_{r}$ is the detuning from $|1\rangle$ to $|r\rangle = |54P_{3/2}, m_J=+3/2\rangle$ transition. (b) Energy level diagram of an ensemble of $N$ bare state atoms, symmetrically coupled under the assumption of a perfect Rydberg blockade; $\ket{g,n} \equiv \big\{ \ket{0}^{\otimes N-n}  \ket{1}^{\otimes n} \big\}_{\rm sym}$ and $\ket{e,n} \equiv \big\{ \ket{0}^{\otimes N-n-1}  \ket{1}^{\otimes n}  \ket{r} \big\}_{\rm sym}$. (c) Energy level diagram of dressed ground state atoms showing the Autler-Townes splitting and exhibiting the nonlinearity of the JC model. The states of $\ket{\widetilde{g},n}$ and $\ket{\widetilde{e},n-1}$ are the ground-like and Rydberg-like dressed states, respectively; $\ket{\widetilde{g},n} \equiv \cos(\theta_n/2) \ket{g,n} + \sin(\theta_n/2) \ket{e,n-1}$ and $\ket{\widetilde{e},n-1} \equiv \cos(\theta_n/2) \ket{e,n-1} - \sin(\theta_n/2) \ket{g,n}$, where $\tan\theta_n = \sqrt{n}\Omega_r/\Delta_r$.}
\label{fig_energy_level}
\end{figure}

Here we study the Rydberg-dressed protocol in which we use a laser to dress one of the clock states of ground-state cesium with an excited Rydberg state~\cite{Keating13, Keating14, Hankin14, Jau15, Keating16}. The resulting Rydberg-dressed states~\cite{Johnson10}, are a superposition of a Rydberg state and a ground clock state allowing both a strong, tunable EDDI and a long coherence time. We measure the resulting JC ladder with two-photon stimulated Raman spectroscopy on the microwave clock transitions in the ground state manifold. While the nonlinearity of the Hamiltonian plays an important role in our previous experiments~\cite{Hankin14, Jau15}, due to advances in the control of our apparatus, we are now able to explore the JC Hamiltonian in complete detail.

In our model~\cite{Keating16}, each atom is described by three states $\{\ket{0},\ket{1},\ket{r}\}$, where we encode in the $^{133}$Cs clock states, $\ket{0} = \ket{6S_{1/2}, F=3,m_F=0}$ and $\ket{1} = \ket{6S_{1/2}, F=4,m_F=0}$,  and $\ket{r}=\ket{54P_{3/2}, m_J=3/2}$ is a chosen Rydberg level coupled to $\ket{1}$ via an optical transition (the detuning is small compared to the clock-state splitting) as shown in Fig.~\ref{fig_energy_level} (a).  The Hamiltonian for the atomic ensemble is $\hat{H}=\sum_{i=1}^N \hat{H}^{(i)} + \hat{V}_{DD}$, where
\begin{align}
\hat{H}^{(i)}&=\hbar \omega_{HF}\ket{1}\bra{1}^{(i)}+(\hbar \omega_{HF}-\hbar \Delta_r)\ket{r}\bra{r}^{(i)} \nonumber \\
&+\frac{\hbar\Omega_r}{2}\left(\ket{r}\bra{1}^{(i)}+\ket{1}\bra{r}^{(i)}\right),
\end{align}
is the Hamiltonian for the $i^{th}$ atom coupling to the laser in the rotating frame, and where $\hbar \omega_{HF}$ is the ground-state hyperfine splitting; zero energy is set at the $\ket{0}$ state. We assume the laser Rabi frequency $\Omega_r$ and detuning $\Delta_r$ are equal for all atoms.  $\hat{V}_{DD}$ is the Hamiltonian describing the EDDI-induced blockade interaction.  If we assume that all atoms are within the blockade radius, and that the blockade is perfect, the spatial dependence of  $\hat{V}_{DD}$ is no longer relevant and the Hamiltonian is symmetric under the exchange of any two atoms. We can thus restrict our attention solely to symmetric states.  For a fixed total number of atoms $N$, the basis for the symmetric subspace is determined by two indices, the number of atoms $n_r$ in the $\ket{r}$ state, and the number of atoms $n$ in the $\ket{1}$ state. In the perfect blockade limit, second-and-higher excitations are prevented, and $\hat{V}_{DD}$ can be conveniently accounted for by projecting into the subspace of states with $n_r\leq 1$. Our state space of interest is thus indexed by $n$ and a binary variable $\{e, g\}$ denoting the presence of a Rydberg excitation or all atoms in the electronic ground state.  Explicitly, the bare basis states for the symmetric, perfectly blockaded subspace are
\begin{equation}
\begin{split}
&\ket{g,n} \equiv \big\{ \ket{0}^{\otimes N-n}  \ket{1}^{\otimes n} \big\}_{\rm sym},\\
&\ket{e,n} \equiv \big\{ \ket{0}^{\otimes N-n-1}  \ket{1}^{\otimes n}  \ket{r} \big\}_{\rm sym},
\end{split}
\end{equation}
where ``sym'' denotes symmetrization over all possible permutations. Note, the states $\ket{g,n}$ are the Dicke states associated with $N$-qubits~\cite{Dicke54}. For example $\ket{g,1}$ is the W-state, associated with one atom excited to $\ket{1}$ and the remainder in $\ket{0}$  (see Fig.~\ref{fig_energy_level} (b)).  

\begin{figure}[t!]
\includegraphics[width=0.7\columnwidth]{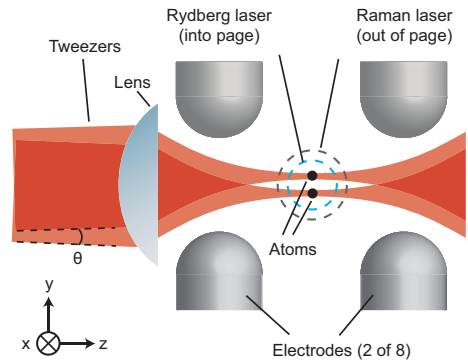}
\caption{Experimental Setup. The Rydberg laser and the Raman lasers are aligned along the x-axis. Two optical tweezers are formed by two lasers with an angular separation $\theta$. In this set-up, eight electrodes null the electric fields near the trapped atoms. The bias magnetic field is applied along the x-axis.}
\label{diagram}
\end{figure}

\begin{figure}[t!]
\includegraphics[width=1\columnwidth]{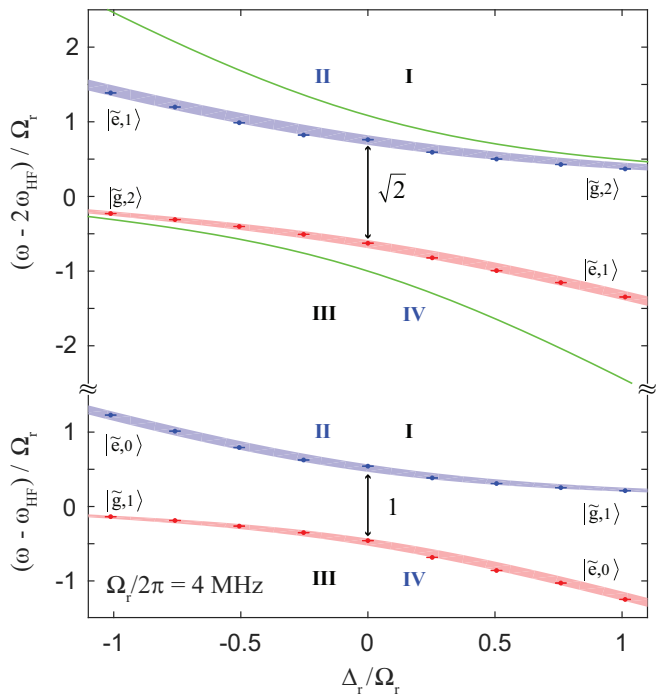}
\caption{Jaynes-Cummings ladder and its $\sqrt{n}$ nonlinearity for two Rydberg-dressed atoms. The x-axis is the normalized Rydberg detuning $\Delta_{r}/\Omega_{r}$, and the y-axis gives the normalized microwave detuning. Two-atom Rydberg-dressed states of $\ket{\widetilde{g},2}$ ($\ket{\widetilde{e},1}$) are located at the upper quadrant I and III (II and IV), and single-atom Rydberg-dressed states of $\ket{\widetilde{g},1}$ ($\ket{\widetilde{e},0}$) are positioned at the lower quadrant I and III (II and IV). The upper/lower quadrant I and III (II and IV) are related to the ground-like (Rydberg-like) dressed states. The red (blue) bands are theoretical predictions of the microwave frequencies for the Autler-Townes splitting, incorporating measured systematic drifts (5~\%) in the experimental parameters ($\Delta_r$ and $\Omega_{r}$). The green lines correspond to the energy of two atoms without the interaction. Two Rydberg transitions are considered for the theoretical plots, $|6 S_{1/2}, F=4, m_{f}=0\rangle$ $\rightarrow$ $|54 P_{3/2}, m_J = +3/2\rangle$ and  $|54 P_{3/2}, m_J = +1/2\rangle$. The Rabi frequencies of $m_J$ = +3/2 and +1/2 are $\Omega_{r}$ and $\Omega_{r}/\sqrt{3}$, respectively. States with $m_J$ = +3/2 and +1/2 are separated by $\Delta_{\rm Zeeman} \simeq 2.13\,\Omega_{r}$ with a 4.6\,G magnetic field.}
\label{fig_AT}
\end{figure}

The Rydberg laser excites atoms in the bare ground state $\ket{g,n}$ to its Rydberg counterpart $\ket{e,n}$ state, with a Rabi frequency enhanced by the collective factor $\sqrt{n}$. The total Hamiltonian in the bare symmetric basis is thus
\begin{equation} \label{eq:almost_JC}
\begin{split}
\hat{H} &= \displaystyle\sum_{n=0}^N \Big(n \hbar \omega_{HF}\ket{g,n}\bra{g,n}\\
&+  (n \hbar \omega_{HF}-\hbar \Delta_r)\ket{e,n-1}\bra{e,n-1}\\
&+  \frac{\sqrt{n} \hbar \Omega_r}{2}\Big(\ket{e,n-1}\bra{g,n}+\ket{g,n}\bra{e,n-1}\Big)\Big).
\end{split}
\end{equation}
This is isomorphic to the well known cavity QED JC Hamiltonian $\hat{H}_{JC} = \hbar \omega_{c} \hat{a}^{\dagger} \hat{a} + \hbar  \omega_{eg}  \hat{\sigma}_{+} \hat{\sigma}_{-} + \hbar g  \left( \hat{a}^{\dagger} \hat{\sigma}_{-} + \hat{a}\hat{\sigma}_{+} \right)$, projected onto the bare basis for up to $N$ excitations~\cite{Keating16}. Here the cavity frequency $\omega_c \rightarrow \omega_{HF}$; the bosonic mode is the number of atoms symmetrically excited to $\ket{1}$. The two-level system is the presence or absence of the symmetric single Rydberg excitation, with energy  $\omega_{eg} \rightarrow \omega_{HF}-\Delta_r$.  The vacuum Rabi frequency $2g \rightarrow \Omega_r$.  The eigenstates of the JC Hamiltonian are the Rydberg-dressed states, 
\begin{align}
\ket{+,n} &= \cos\frac{\theta_n}{2} \ket{g,n} + \sin\frac{\theta_n}{2} \ket{e,n-1} \equiv \ket{\widetilde{g},n},\\
\ket{-,n} &=  \cos\frac{\theta_n}{2} \ket{e,n-1} - \sin\frac{\theta_n}{2} \ket{g,n} \equiv \ket{\widetilde{e},n-1},
\end{align}
where $\tan\theta_n = \sqrt{n}\Omega_r/\Delta_r$. In the case of $\Delta_{r}>0~(<0)$, the ground-like dressed state $\ket{\widetilde{g},n}$ is in the blue (red)-detuned microwave frequencies, and the Rydberg-like dressed state $\ket{\widetilde{e},n-1}$ is in the red (blue)-detuned microwave frequencies as shown in Fig.~\ref{fig_AT}. The JC ladder is given by eigenvalues of the dressed states, 
\begin{equation}
E_{n,\pm} = n\hbar \omega_{HF}+\frac{\hbar}{2}\left(-\Delta_r \pm {\rm sign} (\Delta_r) \sqrt{n \Omega_r^2 + \Delta_r^2}\right),
\end{equation}where the splitting $|E_{n,+}-E_{n,-}|/\hbar = \sqrt{n}\Omega_r$ on Rydberg resonance ($\Delta_r = 0$). As in \cite{Jau15}, this results in an interaction strength between two Rydberg-dressed atoms, which is determined by the nonlinear shift of the dressed state. The nonlinear shifts of the ground-like dressed state ($\kappa_{+}$) and Rydberg-like dressed state ($\kappa_{-}$) are defind as follows:
\begin{align}
\kappa_{+} &= E_{2,+} - 2 E_{1,+} = \langle \widetilde{g},2|\hat{H}|\widetilde{g},2 \rangle - 2 \langle \widetilde{g},1|\hat{H}|\widetilde{g},1 \rangle,\\
\kappa_{-} &= E_{2,-} - 2 E_{1,-} = \langle \widetilde{e},1|\hat{H}|\widetilde{e},1 \rangle - 2 \langle \widetilde{e},0|\hat{H}|\widetilde{e},0 \rangle.
\end{align}
We find \small{$\kappa_{\pm} = \frac{\hbar}{2}\left(\Delta_{r} \pm {\rm sign}(\Delta_{r})(\sqrt{2\Omega_{r}^2 + \Delta_{r}^2} - 2 \sqrt{\Omega_{r}^2 + \Delta_{r}^2})\right)$} (see Fig.~\ref{fig_AT}); the nonlinear shift of the ground-like state $\kappa_{+}$ is defined as $J$ for $\Delta_{r}>0$ in \cite{Jau15}.

We demonstrate this mapping with two Rydberg-dressed cesium atoms. Our apparatus~\cite{Hankin14} employs two laser-cooled $^{133}$Cs atoms confined in optical tweezers at 938\,nm, a magic wavelength for the $^{133}$Cs D2 laser cooling transition. We create the two tweezers by focusing the 938\,nm light that has passed through an AOM (acousto-optic modulator) modulated at two frequencies.  By controlling the two modulation frequencies, we obtain exquisite control over the angular separation $\theta$ of the output and thus the spatial separation of the tweezers as shown in Fig.~\ref{diagram}.  The focusing lens is mounted inside an ultra-high vacuum chamber, avoiding spherical aberrations from vacuum viewports within the focal path.  For detection, atomic fluorescence is collected through this same lens and detected by two single photon detectors. We nullify stray electric fields near the atoms by applying voltages to 8 electrodes surrounding the tweezers (see Fig.~\ref{diagram}).  For fast microwave control ($\sim$1\,MHz) between $\ket{0}$ and $\ket{1}$, we apply two-photon stimulated Doppler-free Raman pulses with a laser detuned $\sim$80\,GHz red from the D2 transition and modulated at the hyperfine frequency with a fiber-based EOM (electro-optic modulator). We couple $\ket{1}$ to Rydberg levels via a 319\,nm single-photon excitation laser~\cite{Hankin14} with Rabi frequencies up to 4.0\,MHz.  This approach offers a good coherence time due to reduced photon scattering, and minimizes dipole forces when compared to a two-photon Rydberg excitation method.

In the experiment, the optically trapped atoms are further cooled to $\sim$10\,$\mu$K using polarization gradient cooling followed by adiabatic lowering of the tweezers potential.  A bias field of 4.6\,G along $\hat{x}$ turns on, and optically pumped atoms to $|1\rangle$ can be detected with $>$ 90\% efficiency. After state preparation, the Cs atoms are brought to a close separation distance of $R \sim 2.84\,\mathrm{\mu m}$ by ramping the AOM modulation frequencies, and a global Raman $\pi$-pulse brings the atoms from $|g, 2\rangle$ to $|g, 0\rangle$. We then apply the Rydberg laser, detuned $\Delta_{r}$ from $\ket{54P_{3/2},m_J=+3/2}$, to dress the $|1\rangle$ states ($|g,n>0\rangle$), and perform spectroscopy by scanning the detuning of a second Raman transition from $|g,0\rangle$ to the Rydberg-dressed states of $\ket{\widetilde{g},2}$, $\ket{\widetilde{e},1}$, $\ket{\widetilde{g},1}$, and $\ket{\widetilde{e},0}$. As in our previous work~\cite{Jau15}, we momentarily extinguish the optical tweezers during this step to avoid additional light shifts from the trapping potential. Following this step, we perform state-dependent detection on each atom to measure the effect of the experiment. We discard events where one or both of the atoms are lost during the experiment and rapidly reuse the same atoms for a subsequent measurement should they both remain. This avoids spoiling of the data by events where either atom projects into the Rydberg state, and enhances the data rate. Compared to our previous work~\cite{Jau15}; nullifying the electric field, controlling optical scatter, and increasing the atom-to-surface distances from $\approx$\,2\,mm to 7\,mm have led to an enhancement of Rydberg state coherence and virtually eliminated Rydberg state loss, leading to a measured lifetime of 116$\pm$19\,$\mu$s.

\begin{figure}[t!]
\includegraphics[width=1\columnwidth]{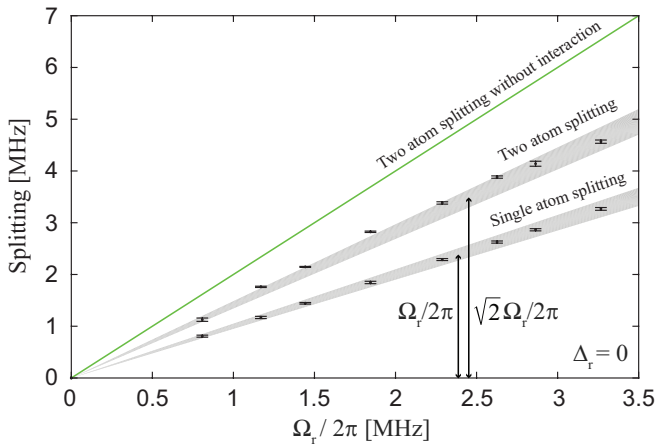}
\caption{Resonant ($\Delta_{r} \approx 0$) Autler-Townes splitting of two Rydberg-dressed atoms as a function of Rydberg-transition Rabi frequency. The x-axis is the Rabi-frequency of a single-atom Rydberg excitation ($\Omega_r$), and the y-axis is the Autler-Townes splitting measured by microwave spectroscopy. The two upper and lower data trends are the splittings of two-atom Rydberg-dressed states between $\ket{\widetilde{g},2}$ and $\ket{\widetilde{e},1}$ and single-atom Rydberg-dressed states between $\ket{\widetilde{g},1}$ and $\ket{\widetilde{e},0}$. The two gray bands incorporate measured systematic drifts (5~\%) in the experimental parameters ($\Omega_{r}$ and $\Delta_{r}$). The green line corresponds to the splitting of two atoms without the interaction. Based on linear fits, the ratio of the splittings is $1.43(0.03)$, consistent with $\sqrt{2}$.}
\label{fig_AT2}
\end{figure}

We observe the JC ladder in the case of $N = 2$. First, the Autler-Townes splitting of Rydberg-dressed ground state atoms is measured by microwave spectroscopy as a function of the detuning of the Rydberg excitation laser as shown in Fig.~\ref{fig_AT}. As the detuning of the Rydberg laser varies, the splitting of the peaks varies according to $\Omega_{eff}=\sqrt{n \Omega_{r}^2 + \Delta_{r}^2}$. We compare this data with theoretical plots that include the effect of both $m_J=+3/2$ and the nearby $m_J=+1/2$ which is necessary to adequately explain the data. Second, we measure the Autler-Townes splitting of two Rydberg-dressed atoms as a function of the Rabi frequency $\Omega_{r}$ for the case of $\Delta_{r}  = 0$ as shown in Fig.~\ref{fig_AT2}.   As expected, the ratio of the slopes between single-atom splitting and two-atom splitting is $1.43(0.03) \approx \sqrt{2}$.

The JC spectrum of Rydberg-dressed states are composed of the ground-like dressed states $\ket{\widetilde{g},n}$ and the Rydberg-like dressed states $\ket{\widetilde{e},n-1}$ determined by $\Delta_{r}$ and $\Omega_{r}$ as shown in Fig.~\ref{fig_AT}. Generally, the Rydberg-like dressed state $\ket{\widetilde{e},1}$ is difficult to measure due to a higher loss rate which reduces the detection efficiency compared to the ground-like dressed state $\ket{\widetilde{g},2}$. Thus we explore the full spectrum of two-atom Rydberg-dressed states with an adiabatic ramping technique. This technique includes ramping $\Delta_{r}$ and $\Omega_{r}$ such that the ensemble transfers from a Rydberg-like dressed state to the bare state $|g, n\rangle$ at the end before projective state measurement. By using this technique, we significantly reduce loss, projection into the Rydberg state, and extend the range of dressed states that we can measure. To detect the dressed states at red (blue)-detuned microwave frequencies (see Fig.~\ref{fig_AT}), opposite frequency ramping directions to map the state back to the ground state are required. The eigenvalues of the dressed states show significantly different spectral character when comparing the ground-like dressed state $\ket{\widetilde{g},2}$ and the Rydberg-like dressed state $\ket{\widetilde{e},1}$. The magnitude of the nonlinear shift of the ground-like dressed states $|\kappa_{+}| = |E_{2,+}-2 E_{1,+}|$ scales as $|\Delta_{r}|^{-3}$ for $|\Delta_{r}| \gg \Omega_{r}$ (see the upper quadrant I and III of Fig.~\ref{fig_AT}). However, the magnitude of the nonlinear shift of the Rydberg-like ground states $|\kappa_{-}| = |E_{2,-} - 2 E_{1,-}|$ scales as $|\Delta_{r}|$ for $|\Delta_{r}| \gg \Omega_{r}$ (see the upper quadrant II and IV of Fig.~\ref{fig_AT}).

In conclusion, we directly observe the full spectrum of the JC ladder and its $\sqrt{n}$ nonlinearity in a two-atom Rydberg-dressed system. The normal-mode splitting of symmetric atomic ensembles with a Rydberg-blockade is the hallmark of the nonlinear coupling of the JC model. Furthermore, the full spectrum of the Rydberg-dressed states could offer a new approach to creating entanglement, operating phase gates, and generating more arbitrary quantum states. Arbitrary symmetric entangled states can be generated using quantum optimal control~\cite{Keating16} or via a collective quantum logic gate using Rydberg superatoms~\cite{Gross15, Ott15}. For example, with highly efficient single-atom loading~\cite{Lester15} and a $\sim$10\,$\mu$m Blockade radius~\cite{Saffman13, Browaeys14}, this could be extended to an ensemble of $\sim$100 atoms in 3D lattices~\cite{Weiss15, Weiss16}.

We also appreciate the assistance of C. H. Baldwin and A. Orozco. This work was supported by the Laboratory Directed Research and Development program at Sandia National Laboratories. Sandia National Laboratories is a multi-mission laboratory managed and operated by Sandia Corporation, a wholly owned subsidiary of Lockheed Martin Corporation, for the U.S. Department of Energy's National Nuclear Security Administration under contract DE-AC04-94AL85000.

\end{document}